\documentclass[twocolumn,aps,prb]{revtex4}

\usepackage{amsmath,amssymb,amsfonts,bm}
\usepackage{graphicx,epstopdf}  %epsfig,
\usepackage{color}
\usepackage{lscape}
\usepackage{extarrows}
\usepackage{enumitem}
\usepackage{empheq}
\usepackage{chngcntr}
\counterwithout{equation}{section}

\allowdisplaybreaks[1]

\newcommand{\w}{\omega}

\newcommand{\B}{\mbox{\tiny B}}

\newcommand{\tS}{\mbox{\tiny S}}
\newcommand{\T}{\mbox{\tiny T}}

\newcommand{\SB}{\mbox{\tiny SB}}

\newcommand{\la}{\langle}
\newcommand{\ra}{\rangle}

\newcommand{\nl}{\nonumber \\}
\newcommand{\be}{\begin{equation}}
\newcommand{\ee}{\end{equation}}
\newcommand{\bsube}{\begin{subequations}}
\newcommand{\esube}{\end{subequations}}
\newcommand{\Eq}[1]{Eq.\,(\ref{#1})}
\newcommand{\Eqs}[1]{Eqs.\,(\ref{#1})}
\newcommand{\Fig}[1]{Fig.\,\ref{#1}}

\newcommand{\RN}[1]{%
  \textup{\uppercase\expandafter{\romannumeral#1}}%
}

\newcommand{\greater}{\mbox{\tiny $>$}}
\newcommand{\lesser}{\mbox{\tiny $<$}}
\newcommand{\lgter}{\mbox{\tiny $\lessgtr$}}

\begin{document}
\title{Entangled system--and--environment dynamics:
Phase--space dissipaton theory}

\author{Yao Wang}
\author{Rui-Xue Xu}   \email{rxxu@ustc.edu.cn}
\author{YiJing Yan}   \email{yanyj@ustc.edu.cn}

\affiliation{Hefei National Laboratory for Physical Sciences at the Microscale
and Department of Chemical Physics
and Synergetic Innovation Center of Quantum Information and Quantum Physics
and
Collaborative Innovation Center of Chemistry for Energy Materials (i{\rm ChEM}),
University of Science and Technology of China, Hefei, Anhui 230026, China}

\date{November 8, 2019}

\begin{abstract}
 Dissipaton--equation--of--motion (DEOM) theory
[Y.~J.~Yan, J.\ Chem.\ Phys.\ {\bf 140}, 054105 (2014)]
is an exact and nonperturbative
many--particle method for open quantum systems.
The existing dissipaton algebra treats also
the dynamics of hybrid bath solvation coordinates.
The dynamics of conjugate momentums remain to be addressed
within the DEOM framework.
In this work, we establish this missing ingredient,
the dissipaton algebra on
solvation momentums,
with rigorous validations against necessary and sufficient criteria.
The resulted phase--space DEOM theory
will serve as a solid ground for further developments
of various practical methods
toward a broad range of applications.
%%%
We illustrate this novel dissipaton algebra with
the phase--space DEOM--evaluation
on heat current fluctuation.

\end{abstract}

\maketitle

 Entangled system--and--environment dynamics play
crucial roles whenever the non-Markovian
and nonperturbative quantum nature of environments
cannot be neglected.
Traditionally this problem was addressed
with the ``core--system'' approach.\cite{Gol002095,Tho012991,Gar854491,Zus80295,Zus8329,Cal83587}
This is to divide the overall environment into the ``first--shell''
and ``secondary'' parts.
%%%
The core-system comprises
both the primary system and the first--shell
hybrid bath solvation modes.
Various approximate theories
such as quantum master equations
had been applied to treat the reduced core-system dynamics
under the influence of the secondary bath environments.%
\cite{Gol002095,Tho012991,Gar854491,Zus80295,Zus8329,Cal83587}

 As exact methods are concerned,
one often exploits the hierarchical--equations--of--motion (HEOM)
formalism.\cite{Tan89101,Tan906676,Tan06082001,Yan04216,Xu05041103,Xu07031107}
%%%
This is the time--derivative equivalence to the
Feynman--Vernon influence functional
path integral formalism,\cite{Fey63118}
which is exact for arbitrary systems coupling
with Gaussian environments.
However, HEOM involves a vast number of dynamical variables
that are just mathematical auxiliaries without
physical meanings.
%%%
In this sense, HEOM is mainly a reduced primary--system theory.
Its exact treatment on the system--plus--solvation
core dynamics is in general too expensive.
%%%%
%{\color{red}
On the other hand, the dissipaton--equation--of--motion
(DEOM) theory,\cite{Yan14054105,Yan16110306,Zha18780}
which recovers the HEOM formalism,
is by far capable of treating the hybrid  bath
solvation coordinate dynamics.
%%%
However, the conjugate momentum
dynamics is yet to be addressed within the DEOM framework.

 The main objective of this work
is to complete the DEOM theory as an exact
and nonperturbative
``core--system'' phase--space method.
To that end, we construct the dissipaton algebra
on the conjugate solvation momentum.
The resulted phase--space DEOM theory
is rigorously validated with respect to
necessary and sufficient criteria.
%
%{\color{magenta}
 It is worth noting that open quantum systems
are beyond the total system--plus--bath
composite Hamiltonian description.
%%%
Additional information, such as temperature $T$
and the interacting bath statistics, would be
needed.\cite{Wei08,Kle09,Yan05187}
The underlying irreversibility will eventually lead to
the total system--plus--bath composite
bulk matter the Boltzmann distribution at a given temperature $T$.
%%%%
In fact, the total composite Hamiltonian
constitutes a ``closed system'' in the thermodynamics nomenclature,
such as a solution system in chemistry,
which is in thermal contact with surroundings.
The solute--and--solvent mixing
free--energy change is dictated by the coupling between them.\cite{Kir35300}
The exact and nonperturbative phase--space DEOM theory
to be developed in this work
would serve as a solid foundation for development of various practical
simulation methods toward realistic molecular systems in condensed phases.
%}%

  Without loss of generality,
we focus explicitly on
a single--dissipative--mode case,
with the system--bath coupling
the form of
$H_{\SB}=\hat Q_{\tS}\hat x_{\B}$.
While the dissipative system mode $\hat Q_{\tS}$
is arbitrary,
the solvation coordinate $\hat x_{\B}$ is linear,
which together with harmonic bath $h_{\B}$
constitute a Gaussian--Wick's environment.
%%%
The total composite Hamiltonian considered
explicitly in this work reads
\be\label{Htotal}
  H_{\T}= H_{\tS}+h_{\B} +{\hat Q}_{\tS}{\hat x}_{\B} .
\ee
Throughout the paper we set $\hbar=1$ and $\beta=1/(k_BT)$, with $k_B$ the Boltzmann constant and $T$ the temperature.
%{\color{magenta}
Let $\rho^{\rm st}_{\T}(H_{\T})$ be
a \emph{steady--state} density operator in the
total system--plus--bath composite space,
which defines the ensemble average,
$\la \,(\cdot)\, \ra
\equiv {\rm Tr}[(\cdot)\rho^{\rm st}_{\T}(H_{\T})]$.
Denote $\hat A(t)\equiv e^{iH_{\T}t}\hat Ae^{-iH_{\T}t}$
for an arbitrary dynamical operator.
Set the time variable $t\geq 0$, unless further specified.
%}
%%

%%%

 Let $\hat x_{\B}$ and $\hat p_{\B}$ be the dimensionless
solvation coordinate and momentum respectively.
The solvation bath frequency $\w_{\B}$
can be specified later in relation
to the Gaussian--Wick's environment description;
see \Eq{omgb2}.
We have
\be\label{xBdot}
 \dot{\hat x}_{\B}=i[H_{\T},\hat x_{\B}]=i[h_{\B},\hat x_{\B}]
=\w_{\B}\hat p_{\B}.
\ee
%{\color{magenta}
This leads to
$\la \dot{\hat A}(t)\hat x_{\B}(0)\ra
=-\la\hat A(t)\dot{\hat x}_{\B}(0)\ra$ the
expression
\be\label{tr1}
 \la \dot{\hat A}(t)\hat x_{\B}(0)\ra=-\w_{\B}\la \hat A(t) {\hat p}_{\B}(0)\ra.
\ee
This together with $[\hat x_{\B},\hat p_{\B}]=i$
constitute the necessary and sufficient requirements
for the to--be--developed dissipaton algebra
on the conjugate solvation momentum,
in addition to the existing DEOM
framework.\cite{Yan14054105,Yan16110306,Zha18780}
%}%

 Let us begin with the common setup
for constructing the DEOM/HEOM theory --
the exponential series expansion
on the interacting bath correlation function via
\be\label{FDT}
\la\hat x^{\B}_{\B}(t)\hat x^{\B}_{\B}(0)\ra_{\B}
=\frac{1}{\pi}
      \!\int^{\infty}_{-\infty}\!\!{\rm d}\omega\,
        \frac{e^{-i\omega t}J_{\B}(\omega)}{1-e^{-\beta\omega}}.
\ee
%{\color{magenta}
This is the fluctuation--dissipation theorem,\cite{Wei08}
with both $\hat x^{\B}_{\B}(t)\equiv e^{ih_{\B}t}\hat x_{\B}e^{-ih_{\B}t}$
and the \emph{equilibrium} ensemble average, $\la(\,\cdot\,)\ra_{\B}
\equiv {\rm tr}_{\B}[(\,\cdot\,)e^{-\beta h_{\B}}]/%
{\rm tr}_{\B}(e^{-\beta h_{\B}})$,
are defined in the uncorrelated bare--bath subspace.
Note that $\hat x^{\B}_{\B}(t)\neq \hat x_{\B}(t)$ and
$\la(\,\cdot\,)\ra_{\B}\neq \la(\,\cdot\,)\ra$ used in \Eq{tr1}.
%}%
The involving hybridization bath spectral density in \Eq{FDT},
which satisfies $J_{\B}(-\w)=-J_{\B}(\w)$,
is given by\cite{Wei08}
\begin{align}\label{JB}
  J_{\B}(\omega)\equiv\frac{1}{2}\!\int^{\infty}_{-\infty}\!\!{\rm d}t\,
    e^{i\omega t} \la[\hat x^{\B}_{\B}(t),\hat x^{\B}_{\B}(0)]\ra_{\B}.
\end{align}
%%%
This also determines the solvation frequency with\cite{Wei08,Yan05187}
\be\label{omgb2}
\w_{\B}
%= \sum_{j}c_j^2\omega_j
=\frac{1}{\pi}\!\int_{-\infty}^{\infty}\!{\rm d}\w\,\w J_{\B}(\w).
\ee
The required  exponential series expansion
on \Eq{FDT} can be achieved by adopting a certain the sum--over--poles scheme
to expand the Fourier integrand there,
followed by Cauchy's contour integration.
Together with the identity $\la\hat x^{\B}_{\B}(0)\hat x^{\B}_{\B}(t)\ra_{\B}
=\la\hat x^{\B}_{\B}(t)\hat x^{\B}_{\B}(0)\ra_{\B}^{\ast}$,
we obtain\cite{Yan16110306}
%%%
\be\label{FBt_corr}
\begin{split}
  &\la\hat x^{\B}_{\B}(t)\hat x^{\B}_{\B}(0)\ra_{\B}
=\sum^K_{k=1}\eta_k e^{-\gamma_k t};
\ \ \, (t\geq 0),
\\ &
\la\hat x^{\B}_{\B}(0)\hat x^{\B}_{\B}(t)\ra_{\B}
  =\sum^{K}_{k=1}\eta_{\bar k}^{\ast} e^{-\gamma_k t};
\ \ \, (\gamma_{\bar k}\equiv \gamma_k^{\ast}).
\end{split}
\ee
The second expression follows the fact that
$\{\gamma_k\}$ must be either real or complex conjugate paired.
The associated index $\bar k\in \{k=1,\cdots,K\}$
is defined via $\gamma_{\bar k}\equiv \gamma_k^{\ast}$.
%%%

%{\color{magenta}
 Note that $\hat x^{\B}_{\B}(0)=\hat x_{\B}(0)=\hat x_{\B}$
and $\dot{\hat x}^{\B}_{\B}(0)=\w_{\B}\hat p_{\B}(0)=\w_{\B}\hat p_{\B}
=\dot{\hat x}_{\B}$, as inferred from \Eq{xBdot}.
%}%
Thus, from \Eq{FDT}, we have
\begin{align}\label{dot_xBcorr}
\la \dot{\hat x}_{\B}{\hat x}_{\B}\ra_{\B}
&=\frac{1}{\pi i}
     \! \int^{\infty}_{-\infty}\!\!{\rm d}\omega\,
        \frac{\w J_{\B}(\omega)}{1-e^{-\beta\omega}}
\nl &=\frac{1}{2\pi i}
     \! \int^{\infty}_{-\infty}\!\!{\rm d}\omega\,
        \w J_{\B}(\omega)
\nl &=\frac{\w_{\B}}{2i}.
\end{align}
While the second expression uses the property of
$J_{\B}(-\omega)=-J_{\B}(\omega)$,
the last one is from \Eq{omgb2}.
Moreover, from the identity,
$\la \dot{\hat x}_{\B}{\hat x}_{\B}\ra_{\B}
 =
-\la \hat x_{\B}\dot{\hat x}_{\B}\ra_{\B}$,
and \Eq{FBt_corr}, we obtain
\be\label{use1}
  \w_{\B}=-2i\sum_{k}^{}\!\gamma_k\eta_k
 =2i\sum_{k}^{}\!\gamma_k\eta_{\bar k}^{\ast}.
\ee
We have also
\be\label{delta_F2}
 \la \hat x^2_{\B}\ra_{\B} = \sum_{k}^{}\eta_k= \sum_{k}^{}\eta_{\bar k}^{\ast}.
\ee

Turn to the existing DEOM framework.%
\cite{Yan14054105,Yan16110306,Zha18780}
%%%
First of all, dissipatons
$\{\hat f_{k}\}$ are statistically independent quasi--particles
and characterize
Gaussian solvation environments.
To reproduce \Eq{FBt_corr}, we set
\be\label{hatFB_in_f}
 \hat x_{\B}=\sum^K_{k=1}  \hat f_{k},
\ee
$\hat f_{k}(t)\equiv e^{ih_{\B}t}\hat f_{k}e^{-ih_{\B}t}$, and
\be\label{ff_corr}
\begin{split}
  \la \hat f_{k}(t)\hat f_{k'}(0)\ra_{\B}
&=\la \hat f_{k}\hat f_{k'} \ra_{\B}^{\greater}\,e^{-\gamma_{k}t}
 =\delta_{k k'}\eta_{k} e^{-\gamma_{k}t},
\\ %\label{ff_corr_nu_rev}
  \la \hat f_{k'}(0)\hat f_{k}(t)\ra_{\B}
&= \la \hat f_{k'}\hat f_{k} \ra_{\B}^{\lesser}\,e^{-\gamma_{k}t}
 =\delta_{k k'} \eta_{\bar k}^{\ast} e^{-\gamma_{k}t},
\end{split}
\ee
with
%{\color{red}
$\la \hat A\hat B\ra^{\greater}\equiv \la \hat A(0+)\hat B(0)\ra$ and
$\la \hat A\hat B\ra^{\lesser}\equiv \la \hat A(0)\hat B(0+)\ra$.
%}
Each forward--and--backward pair of dissipaton correlation functions
are associated with a single--exponent $\gamma_k$.
%%%
The resultant generalized diffusion equation
reads \cite{Yan14054105,Yan16110306}
\begin{align}\label{gendiffu}
   {\rm tr}_{\B}\bigg[\Big(\frac{\partial}{\partial t}\hat f_k\Big)_{\B}\rho_{\T}(t)\bigg]
   =-\gamma_k{\rm tr}_{\B}\big[\hat f_k\rho_{\T}(t)\big].
\end{align}
%%%

 Dynamical variables in DEOM are
the dissipaton--augmented--reduced density operators
(DDOs):\cite{Yan14054105,Yan16110306,Zha18780}
\be \label{DDO}
  \rho^{(n)}_{\bf n}(t)\equiv \rho^{(n)}_{n_1\cdots n_K}(t)
\equiv {\rm tr}_{\B}\big[
  \big(\hat f_{K}^{n_K}\cdots\hat f_{1}^{n_1}\big)^{\circ}\rho_{\T}(t)
 \big].
\ee
%%%
Here, $n=n_1+\cdots+n_{K}$, with $n_k\geq 0$
for bosonic dissipatons.
The product of dissipaton operators inside $(\cdots)^\circ$
is \emph{irreducible}, satisfying
$(\hat f_{k}\hat f_{j})^{\circ}
=(\hat f_{j}\hat f_{k})^{\circ}$
for bosonic dissipatons.
%%%
Each $n$--particles DDO, $\rho^{(n)}_{\bf n}(t)$, is specified with
an ordered set of indexes, ${\bf n}\equiv \{n_1\cdots n_K\}$.
%%%
Denote for later use also ${\bf n}^{\pm}_{k}$ that differs from ${\bf n}$ only
at the specified $\hat f_{k}$-disspaton participation number
$n_{k}$ by $\pm 1$.
The reduced system density operator is just
$\rho_{\bf 0}^{(0)}\equiv \rho_{0\cdots 0}^{(0)}$.

In \Eq{DDO}, $\rho_{\T}(t)$ denotes
the total composite density density operator
that satisfies %the Liouville--von Neumann equation,
\be\label{Sch_eq}
 \dot\rho_{\T}=-i[H_{\T},\rho_{\T}]
 =-i[H_{\tS}+h_{\B}+\hat Q_{\tS}\hat x_{\B},\rho_{\T}].
\ee
For presenting the related dissipaton algebra,
we adopt hereafter the following notations:
\be\label{DDO_action}
\begin{split}
  \rho^{(n)}_{\bf n}(t;\hat A^{\times})\equiv
  {\rm tr}_{\B}\Big[\big(\hat f_{K}^{n_k}\cdots\hat f_{1}^{n_1}\big)^{\circ}
  \hat A^{\times}\rho_{\T}(t)\Big],
\\
 \rho^{(n)}_{\bf n}(t;\hat A^{\lgter})\equiv
  {\rm tr}_{\B}\Big[\big(\hat f_{K}^{n_k}\cdots\hat f_{1}^{n_1}\big)^{\circ}
  \hat A^{\lgter}\rho_{\T}(t)\Big],
\end{split}
\ee
where $\hat A^{\times}\equiv \hat A^{\greater}-\hat A^{\lesser}$, with
\be\label{Algrter}
 \hat A^{\greater}\rho_{\T}(t)\equiv \hat A\rho_{\T}(t),
\ \ \
 \hat A^{\lesser}\rho_{\T}(t)\equiv \rho_{\T}(t)\hat A .
\ee
The generalized diffusion equation (\ref{gendiffu})
is now used together with
$(\frac{\partial}{\partial t}\hat f_k)_{\B} =-i[\hat f_k, h_{\B}]$
and results in \cite{Yan14054105}
\be\label{gendiff}
 \rho^{(n)}_{\bf n}(t;h_{\B}^{\times})=-i\Big(\sum_k n_k \gamma_{k}\Big)\rho^{(n)}_{\bf n}(t).
\ee
This is the dissipaton
algebra on the bath $h_{\B}$--action.

 The action of system--bath coupling is treated with
the generalized Wick's theorem:\cite{Yan14054105,Yan16110306}
\[
\begin{split}
\rho_{\bf n}^{(n)}(t;\hat f_k^{\greater})
&= \rho_{{\bf n}^{+}_{k}}^{(n+1)}(t)
 +\sum_{k'}n_{k'} \la \hat f_{k'}\hat f_{k} \ra_{\B}^{\greater}
 \rho_{{\bf n}^{-}_{k'}}^{(n-1)}(t),
\\
\rho_{\bf n}^{(n)}(t;\hat f_k^{\lesser})
&= \rho_{{\bf n}^{+}_{k}}^{(n+1)}(t)
 +\sum_{k'}n_{k'} \la \hat f_{k}\hat f_{k'} \ra_{\B}^{\lesser}
 \rho_{{\bf n}^{-}_{k'}}^{(n-1)}(t).
\end{split}
\]
%{\color{magenta}
Here, $\la \hat f_{k}\hat f_{k'} \ra_{\B}^{\greater}
 =\delta_{k k'}\eta_{k}$
and $\la \hat f_{k}\hat f_{k'} \ra_{\B}^{\lesser}
 =\delta_{k k'}\eta^{\ast}_{\bar k}$ via \Eq{ff_corr}.
Together with \Eq{hatFB_in_f}, we have
\be\label{wick_x}
\rho^{(n)}_{\bf n}(t;\hat x_{\B}^{\lgter})
=\sum_{k}\rho^{(n)}_{\bf n}(t;\hat f_k^{\lgter}),
\ee
%}%
with
\be\label{genwicks}
\begin{split}
\rho^{(n)}_{\bf n}(t;\hat f_k^{\greater})
&= \rho_{{\bf n}^{+}_{k}}^{(n+1)}(t)
 +n_{ k} \eta_{ k}\rho_{{\bf n}^{-}_{ k}}^{(n-1)}(t),
\\
\rho^{(n)}_{\bf n}(t;\hat f_k^{\lesser})
&= \rho_{{\bf n}^{+}_{k}}^{(n+1)}(t)
 +n_{ k} \eta^{\ast}_{ \bar k}\rho_{{\bf n}^{-}_{ k}}^{(n-1)}(t) .
\end{split}
\ee
By applying \Eq{Sch_eq} for the total
composite density operator $\rho_{\T}(t)$ in \Eq{DDO},
followed by using \Eqs{gendiff}--(\ref{genwicks}),
we obtain \cite{Yan14054105}
\begin{align}\label{DEOM}
 \dot\rho^{(n)}_{\bf n}=
& -\Big(iH^{\times}_{\tS}+\sum_k n_k \gamma_{k}\Big)\rho^{(n)}_{\bf n}
  -i\sum_{k}\hat Q^{\times}_{\tS}\rho^{(n+1)}_{{\bf n}_{k}^+}
\nl&
  -i\sum_{k}n_{k}\big(\eta_{k}\hat Q^{\greater}_{\tS}
   -\eta_{\bar k}^{\ast}\hat Q^{\lesser}_{\tS}\big)
  \rho^{(n-1)}_{{\bf n}_{k}^-}.
\end{align}
%{\color{magenta}
As this recovers the HEOM formalism,%
\cite{Tan89101,Tan906676,Tan06082001,Yan04216,Xu05041103,Xu07031107}
we would have also \emph{de facto} validated the existing
dissipaton algebra on the solvation coordinate $\hat x_{\B}$,
\Eqs{hatFB_in_f}--(\ref{genwicks}).
%}%

%{\color{magenta}
 We are now in the position to develop
the dissipaton algebra on the
solvation momentum $\hat p_{\B}$,
in addition to the existing DEOM framework.
%}%
To proceed, we set in analogy to \Eq{hatFB_in_f} the decomposition,
\be\label{pinphi}
  {\hat p}_{\B}=\sum^{K}_{k=1}\hat \varphi_{k},
\ee
thus
\be\label{wick_p}
\rho^{(n)}_{\bf n}(t;\hat p_{\B}^{\lgter})
=\sum_{k}\rho^{(n)}_{\bf n}(t;\hat \varphi_k^{\lgter}).
\ee
%%%
The proposed new ingredients read
\be\label{genwicks_momentum}
\begin{split}
\rho^{(n)}_{\bf n}(t;\hat \varphi_k^{\greater})
&=-\frac{\gamma_k}{\w_{\B}}\big[
 \rho^{(n+1)}_{{\bf n}^{+}_{k}}(t)
 -n_k\eta_k\rho^{(n-1)}_{{\bf n}^{-}_{k}}(t)\big],
\\
\rho^{(n)}_{\bf n}(t;\hat \varphi_k^{\lesser})
&=-\frac{\gamma_k}{\w_{\B}}\big[
 \rho^{(n+1)}_{{\bf n}^{+}_{k}}(t)
 -n_k\eta_{\bar k}^{\ast}\rho^{(n-1)}_{{\bf n}^{-}_{k}}(t)\big].
\end{split}
\ee
The proof below goes with their satisfying
two basic requirements.

Consider the result on
%the requirement One is concerned with
 $[\hat x_{\B}, {\hat p}_{\B}]=i$.
In the DEOM--space algebra, this necessary requirement amounts to

\be\label{canonical}
\rho_{\bf n}^{(n)}\big(t;[\hat x_{\B}, {\hat p}_{\B}]^{\lgter}\big)
=i\rho_{\bf n}^{(n)}(t).
\ee
%%%
\emph{Proof} -- Equation (\ref{genwicks_momentum}),
together with \Eq{genwicks}, lead to
\bsube\label{}
\begin{align}
 \rho^{(n)}_{\bf n}\big[t;(\hat f_k{\hat \varphi}_k)^{\greater}\big]
&=\rho^{(n+1)}_{{\bf n}^{+}_{k}}\big(t;{\hat \varphi}_k^{\greater}\big)
  +n_{k}\eta_{k}\rho^{(n-1)}_{{\bf n}^{-}_{k}}\big(t;{\hat \varphi}_k^{\greater}\big)
\nl&
=-\frac{\gamma_k}{\w_{\B}}\Big[
   \rho^{(n+2)}_{{\bf n}^{++}_{kk}}(t)
  -2{n_k\choose 2}\eta_{ k}^2\rho^{(n-2)}_{{\bf n}^{--}_{ kk}}(t)\Big]
\nl&\quad
  +\frac{\gamma_k\eta_{k}}{\w_{\B}}\rho^{(n)}_{{\bf n}}(t),
%%%%
\intertext{and}
 \rho^{(n)}_{\bf n}\big[t;( {\hat \varphi}_{ k}\hat f_k)^{\greater}\big]
&=-\frac{\gamma_k}{\w_{\B}}\big[
   \rho^{(n+1)}_{{\bf n}^{+}_{k}}(t; {\hat f}_{ k}^{\greater})
  -n_{k}\eta_{k}\rho^{(n-1)}_{{\bf n}^{-}_{k}}(t;{\hat f}_{k}^{\greater})
 \big]
\nl&
=-\frac{\gamma_k}{\w_{\B}}\Big[\rho^{(n+2)}_{{\bf n}^{++}_{kk}}(t)
 -2{n_k\choose 2}\eta_{k}^2\rho^{(n-2)}_{{\bf n}^{--}_{ kk}}(t)\Big]
\nl&\quad
 -\frac{\gamma_k\eta_{k}}{\w_{\B}}\rho^{(n)}_{{\bf n}}(t).
\end{align}
\esube
%{\color{red}
The
$\rho^{(n)}_{\bf n}\big[t;(\hat f_k{\hat \varphi}_k)^{\lesser}\big]$
and $\rho^{(n)}_{\bf n}\big[t;( {\hat \varphi}_{ k}\hat f_k)^{\lesser}\big]$ counterparts
are similar to
$\rho^{(n)}_{\bf n}\big[t;({\hat\varphi}_k\hat f_k)^{\greater}\big]$
and $\rho^{(n)}_{\bf n}\big[t;(\hat f_k{\hat\varphi}_{k})^{\greater}\big]$,
respectively, but with $\eta_{k}$ being replaced by
$\eta^{\ast}_{\bar k}$.
%%%
Together with
\[ %\be\label{cc1}
\rho^{(n)}_{\bf n}\big(t;[\hat f_k,  {\hat \varphi}_{ k}]^{\lgter}\big)
\!\equiv\! \rho^{(n)}_{\bf n}
 \big[t;(\hat f_k{\hat \varphi}_k)^{\lgter}\big]
 -\rho^{(n)}_{\bf n}
  \big[t;({\hat \varphi}_k\hat f_k)^{\lgter}\big],
\]
we obtain
\be \label{fphicommute}
\begin{split}
\rho^{(n)}_{\bf n}\big(t;[\hat f_k,{\hat \varphi}_{k'}]^{\greater}\big)
&=2\delta_{kk'}\frac{\gamma_{k}\eta_{k}}{\w_{\B}}\rho^{(n)}_{\bf n}(t),
\\
\rho^{(n)}_{\bf n}\big(t;[\hat f_k,  {\hat \varphi}_{k'}]^{\lesser}\big)
&=-2\delta_{kk'}\frac{\gamma_{k}\eta^\ast_{\bar k}}{\w_{\B}}
 \rho^{(n)}_{\bf n}(t).
\end{split}
\ee
%\ee
Consequently,
\be\label{temp}
\begin{split}
\rho_{\bf n}^{(n)}\big(t;[\hat x_{\B}, {\hat p}_{\B}]^{\greater}\big)
&=\frac{2}{\w_{\B}}
 \Big(\sum_{k}\gamma_{k}\eta_{k}\Big)\rho^{(n)}_{\bf n}(t),
\\
\rho_{\bf n}^{(n)}\big(t;[\hat x_{\B}, {\hat p}_{\B}]^{\lesser}\big)
&=-\frac{2}{\w_{\B}}\Big(\sum_{k}\gamma_{k}\eta^\ast_{\bar k}\Big)
  \rho^{(n)}_{\bf n}(t).
\end{split}
\ee
%}
Together with the expression of $\w_{\B}$ of \Eq{use1},
we obtain \Eq{canonical}.
%{\color{red}
 It is worth noting that
\Eq{fphicommute} would imply that
$[\hat f_k,  {\hat \varphi}_{k'}]^{\greater}
=2\delta_{kk'}\gamma_{k}\eta_{ k}/\w_{\B}$
%%%
and $[\hat f_k,  {\hat \varphi}_{k'}]^{\lesser}
=2\delta_{kk'}\gamma_{k}\eta^{\ast}_{\bar k}/\w_{\B}$.
%%%
The information on $[\hat f_k,  {\hat \varphi}_{k}]$ is undetermined.
However, the necessary requirement is concerned only with
$\sum_{k}[\hat f_k,  {\hat \varphi}_{k}]
=[\hat x_{\B},\hat p_{\B}]=i$.
Its DEOM--space equivalence, \Eq{canonical},
has just been proved to be satisfied by
the new ingredients of dissipaton algebra
in \Eq{genwicks_momentum}.
%}%

\begin{figure}%[!htbp]
\includegraphics[width=0.48\textwidth]{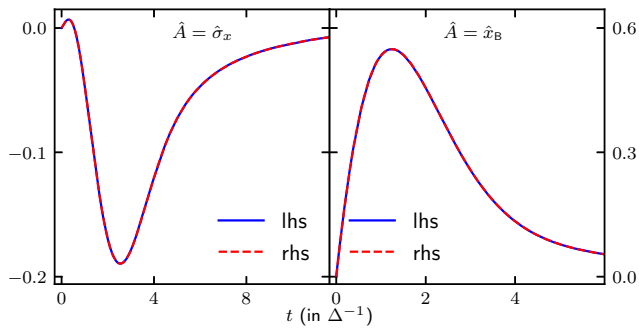}
\caption{Numerical validation of \Eq{genwicks_momentum}  via the equality between lhs\ and rhs\ of \Eq{tr1},
exemplified with $\hat A=\hat\sigma_{x}$ (left panel) and $\hat x_{\B}$ (right panel) for $ H_{\tS}=\frac{\Delta}{2}\hat \sigma_z+\frac{V}{2}\hat \sigma_x$ and $\hat Q_{\tS}=\Delta\hat \sigma_x$.  Bath spectral density is $J_{\B}(\w)=(\w_{\B}\zeta_{\B}\w)/[(\w_{\B}^2-\w^2)^2+\zeta_{\B}^2\w^2]$. Parameters are $V=\w_{\B}=\zeta_{\B}=k_BT=\Delta$. Plotted are only the real parts, whereas the imaginary parts are also equal (not shown here).
}\label{fig1}
\end{figure}

%{\color{red}
 The sufficient requirement is concerned
with \Eq{tr1}, since it engages
an arbitrary dynamical operator $\hat A$.
We exploit \Eq{genwicks_momentum}
to evaluate $\la \hat A(t) {\hat p}_{\B}(0)\ra$
in the right-hand-side (rhs)
of \Eq{tr1}. In parallel, we evaluate
the left-hand-side (lhs) of \Eq{tr1},
$\la \dot{\hat A}(t)\hat x_{\B}(0)\ra$,
with the well--established dissipaton algebra on
solvation coordinate $\hat x_{\B}$.
The agreement between these two DEOM evaluations
is rigorously proved in Supplymentary Material (SM),\cite{Wan20Supl} on the basis of quantum mechanics linear--space constructions.
We have thus analytically validated \Eq{genwicks_momentum}.
%}%
In \Fig{fig1}, we present
the numerical demonstrations
on the required consistency
with two representing types of $\hat A$, the system--type versus hybrid bath--type.

To conclude this paper,
%{\color{magenta}
we illustrate the novel phase--space dissipaton algebra with the heat current fluctuation.
To this end, we adopt the heat current operator $\hat J$ the convention \cite{He18195437,Son17064308} in analogy to that of electron transport,
\be\label{def_J}
\hat J\equiv-\frac{{\rm d}h_{\B}}{{\rm d}t}=-i[H_{\SB},h_{\B}]=\hat Q_{\tS}\w_{\B}\hat p_{\B}.
\ee
It is noticed there are other conventions of heat current adopted in various contexts.%
\cite{Son17064308,Esp15235440,Sch15224303,
Mot17053013,Mot18113020,Wie19Arxiv1903_11368}
Nevertheless, for the main purpose of the present work, we adopt  \Eq{def_J} to ``visualize'' the dissipaton momentums and the resulted phase--space DEOM--based evaluations.
%}%
\begin{figure}%[!htbp]
\includegraphics[width=0.48\textwidth]{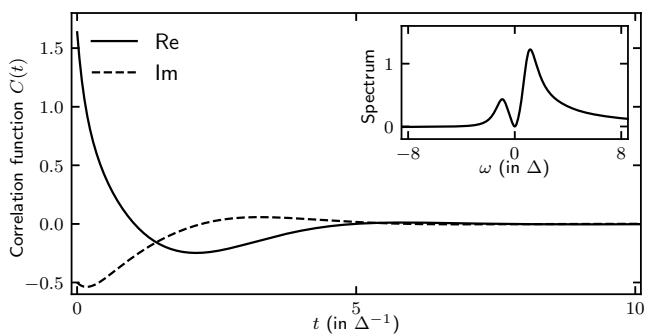}
\caption{Correlation function $C(t)$, \Eq{ct},  together with its spectrum, for the same system of \Fig{fig1}. }\label{fig2}
\end{figure}
Figure {\ref{fig2}} depicts the heat current correlation,
\be\label{ct}
C(t)\equiv\la \delta \hat J(t)\delta\hat J(0)\ra,
\ee
with $\delta \hat J\equiv \hat J-\la\hat J\ra$,
where the average current $\la \hat J \ra=0$ at the equilibrium scenario considered here.
The inset reports the zero--current fluctuation spectrum.
%{\color{red} 
The phase--space DEOM--based quantum mechanics evaluation is detailed in  SM.\cite{Wan20Supl}
%}

In summary, we complete the phase--space DEOM theory with adding the   dissipaton algebra for the hybrid bath solvation momentum.
The new ingredients, \Eqs{pinphi}-(\ref{wick_p}), are shown to satisfy the requirements of both necessary and sufficient criteria.
The exemplified  quantum heat current fluctuation, \Eq{ct} with \Eq{def_J}, is a new class of properties which would not be accessible without the present development.
The resulted phase--space DEOM is a universal and versatile tool
for accurate evaluations on system--plus--hybrid bath dynamics.
%{\color{magenta}
The extension to non-equilibrium scenario involving multiple bath reservoirs and multiple dissipative modes would be straightforward.
It would also be anticipated that the phase--space DEOM theory will serve as a solid ground for development of various
practical simulation methods toward  realistic molecular systems in condensed phases.
%}

%%
See supplementary materials including the DEOM--space quantum mechanics  and an
analytical proof of dissipaton momentum algebra satisfying \Eq{tr1}.

 Support from
the Ministry of Science and Technology of China (Nos.\ 2017YFA0204904 \&
2016YFA0400904),
the Natural Science Foundation of China (Nos.\ 21633006 \& 21703225)
and Anhui Initiative in Quantum Information Technologies
is gratefully acknowledged.

%\bibliographystyle{../aiptit}
%\bibliography{../bibrefs}

\end{document}

\title{Supplementary materials on ``Entangled system--and--environment dynamics:
Phase--space dissipaton theory''}

\author{Yao Wang}
\author{Rui-Xue Xu}   \email{rxxu@ustc.edu.cn}
\author{YiJing Yan}   \email{yanyj@ustc.edu.cn}

\affiliation{Hefei National Laboratory for Physical Sciences at the Microscale
and Department of Chemical Physics
and Synergetic Innovation Center of Quantum Information and Quantum Physics
and
Collaborative Innovation Center of Chemistry for Energy Materials (i{\rm ChEM}),
University of Science and Technology of China, Hefei, Anhui 230026, China}

\date{November 8, 2019}

\begin{abstract}
These supplementary materials on ``Entangled system--and--environment dynamics:
Phase--space dissipaton theory'' contain  (1) A comprehensive description of DEOM--space quantum mechanics, which had been well--established;
(2) An analytical proof of the fact that the dissipaton momentum
as Eq.\,(24) algebra satisfies Eq.\,(3).
\end{abstract}

\maketitle

\section{DEOM--space quantum mechanics}
The DEOM theory describes both the reduced system and hybrid bath dynamics.
 For later convenience, we repeat the Eq.\,(21):
\begin{align}\label{deom}
 \dot\rho^{(n)}_{\bf n}=
& -\Big(iH^{\times}_{\tS}+\sum_k n_k \gamma_{k}\Big)\rho^{(n)}_{\bf n}
  -i\sum_{k}\hat Q^{\times}_{\tS}\rho^{(n+1)}_{{\bf n}_{k}^+}
%\nl&
  -i\sum_{k}n_{k}(\eta_{k}\hat Q^{\greater}_{\tS}
   -\eta_{\bar k}^{\ast}\hat Q^{\lesser}_{\tS})
  \rho^{(n-1)}_{{\bf n}_{k}^-}.
\end{align}
It is worth re-emphasizing that the DEOM theory contains not only the above hierarchical dynamics, but also the dissipaton algebra. In particular, this includes  Eq.\,(19) with Eq.\,(20), and also  Eq.\,(23) with Eq.\,(24) that is the new ingredient developed in the present work.
As results,
the DEOM theory now
supports the accurate evaluations on the expectation values and correlation functions engaging operators of the following types,
\be\label{adeom}
\hat A\in \{\hat A_{\tS}, \hat A_{\tS}\hat x_{\B}, \hat A_{\tS}\hat p_{\B}\},
\ee
with individual $\hat A_{\tS}$ being  an arbitrary system dynamical operator.
\subsection{Expectation values via the DEOM--space mapping}
%%%
The DEOM--space quantum mechanics is a natural extension of Liouville--space formalism.
To this end, we express the \Eq{deom} in the form of
\be\label{sdeom}
\dot{\bm\rho}(t)=-i\bm{\mathcal{L}}{\bm\rho}(t).
\ee
This resembles $\dot{\rho}_{\T}=-i{\cal L}_{\T}\rho_{\T}$ with mapping the total system--plus--bath composite  Liouvillian to  the DEOM--space dynamics generator as defined in
\Eq{deom}, ${\cal L}_{\T}\rightarrow \bm{\mathcal{L}}$,
and
\be
\rho_{\T}(t)\rightarrow {\bm \rho}(t)=\{\rho_{\bf n}^{(n)}(t); n=0,1,2,\cdots\}.
\ee
In parallel,
we should map dynamical operators  into the DEOM--space,
\be\label{mapping}
\hat A \rightarrow \hat {\bm A}\equiv\{\hat A_{\bf n}^{(n)}; n=0,1,2,\cdots\},
\ee
such that the expectation value can be evaluated via
\be\label{expectation}
\bar A(t)\equiv {\rm Tr}[\hat A \rho_{\T}(t)]\equiv\lla\hat A|\rho_{\T}(t) \rra =\lla\hat {\bm A}|{\bm\rho}(t) \rra\equiv \sum_{\bf n}{\rm tr}_{\tS}\big[\hat A_{\bf n}^{(n)}\rho_{\bf n}^{(n)}(t)\big].
\ee
The DEOM--space correspondences to the three types of dynamical operators
in \Eq{adeom} are
\begin{align}\label{smap}
\hat A_{\tS}&\rightarrow \hat {\bm A}=\{{\hat A^{(0)}=\hat A_{\tS}; \ \ \hat A_{\bf n}^{(n>0)}=0}\},
\\
\label{sxmap}
\hat A_{\tS}\hat x_{\B}&\rightarrow \hat {\bm A}=\{\,{\hat A^{(1)}_{k}=\hat A_{\tS}; \ \ \hat A_{\bf n}^{(n\neq 1)}=0}\},
\\  \label{spmap}
\hat A_{\tS}\hat p_{\B}&\rightarrow \hat {\bm A}=\{\,{\hat A^{(1)}_{k}=-(\gamma_k/\w_{\B})\hat A_{\tS}; \ \ \hat A_{\bf n}^{(n\neq 1)}=0}\}.
\end{align}
%The required DEOM--space mapping, \Eq{mapping}, can be readily established based on the generalized Wick's theorem [cf. Eqs.(19)--(20) and Eqs.(23)--(24)].
The first two relations had been established in our previous works. Demonstrated properties of study include
such as Fano interference,\cite{Zha15024112}
Herzberg--Teller vibronic coupling spectroscopy,\cite{Zha16204109}
and transport current shot noise spectrum.\cite{Jin15234108}
The
relation (\ref{spmap}) can be obtain as follows.
According to \Eq{expectation},
\be
{\rm Tr}[\hat A_{\tS}\hat p_{\B}\rho_{\T}(t)]={\rm tr}_{\tS}\big\{\hat A_{\tS}{\rm tr}_{\B}[\hat p_{\B}\rho_{\T}(t)]\big\}={\rm tr}_{\tS}\big\{\hat A_{\tS}\rho_{\bf 0}^{(0)}(t;\hat p_{\B}^{\greater})\big\}=-\sum_{k}(\gamma_k/\w_{\B}){\rm tr}_{\tS}\big\{\hat A_{\tS}\rho_{k}^{(1)}(t)\big\}.
\ee
The last identity that leads to \Eq{spmap} is obtained  by using Eqs.\,(23) and (24).

\subsection{Correlation functions via the DEOM--space mapping}
Mathematically, one can recast correlation functions in form of  expectation values as [cf.\,\Eq{expectation}]
\be\label{corr_hilbert}
\la \hat A(t) \hat B(0)\ra=\lla \hat A|e^{-i{\cal L}_{\T}t}\hat B^{\greater}|\rho_{\T}^{\rm st}\rra\equiv\lla \hat A|\rho_{\T}(t;\hat B^{\greater})\rra
\ee
where
\be
\rho_{\T}(t;\hat B^{\greater})\equiv e^{-i{\cal L}_{\T}t}(\hat B^{\greater}\rho_{\T}^{\rm st})
\equiv
e^{-i{\cal L}_{\T}t}(\hat B\rho_{\T}^{\rm st}).
\ee
In line with \Eq{expectation}, we obtain
\be\label{corr_deom}
\la \hat A(t) \hat B(0)\ra=\lla \hat {\bm A}|{\bm\rho}(t;\hat B^{\greater})\rra=\sum_{\bf n}{\rm tr}_{\tS}\big[\hat A_{\bf n}^{(n)}\rho_{\bf n}^{(n)}(t;\hat B^{\greater})\big].
\ee
Here,
$
{\bm\rho}(t;\hat B^{\greater})\equiv \{\rho_{\bf n}^{(n)}(t;\hat B^{\greater})\}
$
satisfies the \Eq{deom}.
The initial values that correspond to
$\rho_{\T}(t=0;\hat B^{\greater})=\hat B^{\greater}\rho_{\T}^{\rm st}$
following Eq.\,(16), is to be identified via
\be
\rho^{(n)}_{\bf n}(0;\hat B^{\greater})\equiv
  {\rm tr}_{\B}\Big[\big(\hat f_{K}^{n_k}\cdots\hat f_{1}^{n_1}\big)^{\circ}
  \hat B^{\greater}\rho^{\rm st}_{\T}\Big].
\ee
We obtained immediately
\be
\begin{split}\label{action}
\rho^{(n)}_{\bf n}(0;\hat B_{\tS}^{\greater})&
=\hat B_{\tS}\rho^{(n);{\rm st}}_{\bf n},
  \\
  \rho^{(n)}_{\bf n}(0; (\hat B_{\tS}\hat x_{\B})^{\greater})&=\hat B_{\tS}\rho^{(n)}_{\bf n}(0;\hat x_{\B}^{\greater}),
  \\
  \rho^{(n)}_{\bf n}(0; (\hat B_{\tS}\hat p_{\B})^{\greater})&=\hat B_{\tS}\rho^{(n)}_{\bf n}(0;\hat p_{\B}^{\greater}),
\end{split}
\ee
for the three types of dynamical operators in \Eq{adeom}, respectively.
From Eq.\,(19) with Eq.\,(20), we obtain
\be\label{wick_x}
\rho^{(n)}_{\bf n}(0;\hat x_{\B}^{\greater})
=\sum_{k}\left[\rho_{{\bf n}^{+}_{k}}^{(n+1);{\rm st}}
 +n_{ k} \eta_{ k}\rho_{{\bf n}^{-}_{ k}}^{(n-1);{\rm st}}\right].
\ee
Moreover, from Eq.\,(23) with Eq.\,(24) we have
\be\label{wick_p}
\rho^{(n)}_{\bf n}(0;\hat p_{\B}^{\greater})
=-\frac{1}{\w_{\B}}\sum_{k}\gamma_k\left[
 \rho^{(n+1);{\rm st}}_{{\bf n}^{+}_{k}}
 -n_k\eta_k\rho^{(n-1);{\rm st}}_{{\bf n}^{-}_{k}}\right].
\ee
The involved $\{\rho_{\bf n}^{(n);{\rm st}}\}$ are the steady-state solutions of \Eq{deom}.

We summarize the DEOM--evaluation on correlation functions as follows.
(\emph{i}) Compute the steady--state DDOs $\{\rho_{\bf n}^{(n);{\rm st}}\}$ via such as the self-consistent iteration method; \cite{Zha17044105}
(\emph{ii}) Determine the  initial values $\{\rho^{(n)}_{\bf n}(0;\hat B^{\greater})\}$ via \Eqs{action}--(\ref{wick_p});
(\emph{iii}) Propagate DDOs with \Eq{deom} to obtain the required $\{\rho^{(n)}_{\bf n}(t;\hat B^{\greater})\}$ for \Eq{corr_deom};
(\emph{iv}) Evaluate $\la \hat A(t) \hat B(0)\ra$ as the expectation value problem by using \Eqs{expectation}--(\ref{spmap}).
Individual $\hat A$ or $\hat B$ can be any dynamical operator in \Eq{adeom}.

\section{Dissipaton momentum algebra  satisfying Eq.(3): Analytical proof}
For convenience, we repeat Eq.(3) here,
\begin{align} \label{tr1}
\la \dot{\hat A}(t)\hat x_{\B}(0)\ra=-\w_{\B}\la \hat A(t) {\hat p}_{\B}(0)\ra.
\end{align}
%Apparently, \Eq{tr1} can be recast as
%\be\label{check2}
%\la \hat A(t)(-i{\cal L}_{\T})\hat x_{\B}(0)\ra=
%-\w_{\B}\la \hat A(t) {\hat p}_{\B}(0)\ra.
%\ee
In the DEOM space [cf. \Eqs{corr_hilbert} and (\ref{corr_deom})], it reads
\be
\lla \hat {\bm A}|\dot{\bm\rho}(t;\hat x_{\B}^{\greater})\rra=-\w_{\B}\lla \hat {\bm A}|{\bm\rho}(t;\hat p_{\B}^{\greater})\rra.
\ee
%%
%Therefore, to prove the dissipaton algebra on solvation momentum, Eq.\,(23) via Eq.\,(24), satisfies \Eq{tr1}, it is sufficient to show that
As $\hat A$ is arbitrary, the above expression amounts to
\be\label{tmp0}
\dot{\bm\rho}(t;\hat x_{\B}^{\greater})=-\w_{\B}{\bm\rho}(t;\hat p_{\B}^{\greater}).
\ee
The proof is as follows.
Let us start with ${\bm\rho}(t;\hat x_{\B}^{\greater})=e^{-i\bm{\mathcal{L}}t}{\bm\rho}(0;\hat x_{\B}^{\greater})$, which leads to
\be
\dot {\bm\rho}(t;\hat x_{\B}^{\greater})= -e^{-i\bm{\mathcal{L}}t}\big[i\bm{\mathcal{L}}{\bm\rho}(0;\hat x_{\B}^{\greater})\big]=e^{-i\bm{\mathcal{L}}t}\dot {\bm\rho}(0;\hat x_{\B}^{\greater}).
\ee
On the other hand, ${\bm\rho}(t;\hat p_{\B}^{\greater})=e^{-i\bm{\mathcal{L}}t}{\bm\rho}(0;\hat p_{\B}^{\greater})$.
Therefore, \Eq{tmp0} is equivalent to $\dot {\bm\rho}(0;\hat x_{\B}^{\greater})=-\w_{\B}{\bm\rho}(0;\hat p_{\B}^{\greater})$. In other words, it is sufficient to prove
\be\label{tmp}
\dot{\rho}_{\bf n}^{(n)}(0;\hat x_{\B}^{\greater})=-\w_{\B}{\rho}_{\bf n}^{(n)}(0;\hat p_{\B}^{\greater}).
\ee
Equations (\ref{deom}) and (\ref{wick_x}) lead to the lhs the expressions,
\begin{align}\label{tmp2}
\dot{\rho}_{\bf n}^{(n)}(0;\hat x_{\B}^{\greater})&=-\Big(iH^{\times}_{\tS}+\sum_k n_k \gamma_{k}\Big)\rho_{\bf n}^{(n)}(0;\hat x_{\B}^{\greater})
  -i\sum_{k}\hat Q^{\times}_{\tS}\rho_{\bf n^{+}_{k}}^{(n+1)}(0;\hat x_{\B}^{\greater})
%\nl&
  -i\sum_{k}n_{k}(\eta_{k}\hat Q^{\greater}_{\tS}
   -\eta_{\bar k}^{\ast}\hat Q^{\lesser}_{\tS})
  \rho^{(n-1)}_{{\bf n}_{k}^-}(0;\hat x_{\B})
  \nl &
  =-\Big(iH^{\times}_{\tS}+\sum_k n_k \gamma_{k}\Big)\sum_{j}\left[\rho_{{\bf n}^{+}_{j}}^{(n+1);{\rm st}}
 +n_{ j} \eta_{ j}\rho_{{\bf n}^{-}_{ j}}^{(n-1);{\rm st}}\right]
  -i\sum_{k,j}\hat Q^{\times}_{\tS}\left[\rho_{{\bf n}^{++}_{kj}}^{(n+2);{\rm st}}
 +(n_{ j}+\delta_{kj}) \eta_{ j}\rho_{{\bf n}^{+-}_{ kj}}^{(n);{\rm st}}\right]
\nl&\quad
  -i\sum_{k,j}n_{k}(\eta_{k}\hat Q^{\greater}_{\tS}
   -\eta_{\bar k}^{\ast}\hat Q^{\lesser}_{\tS})
\left[\rho_{{\bf n}^{-+}_{kj}}^{(n);{\rm st}}
 +(n_{ j}-\delta_{kj}) \eta_{ j}\rho_{{\bf n}^{--}_{ kj}}^{(n-2);{\rm st}}\right].
\end{align}
On the other hand, the steady--state solutions  to \Eq{deom} satisfy (noting that ${\bf n}^{\pm\pm}_{jk}={\bf n}^{\pm\pm}_{kj}$)
\be\label{tmp3}
\begin{split}
&0=-\Big[iH^{\times}_{\tS}+\sum_k (n_k+ \delta_{kj}) \gamma_{k}\Big]\rho^{(n+ 1);{\rm st}}_{{\bf n}^{+}_{j}}
  -i\sum_{k}\hat Q^{\times}_{\tS}\rho_{{\bf n}^{++}_{kj}}^{(n+2);{\rm st}}
%\nl&
  -i\sum_{k}(n_{k}+\delta_{kj})(\eta_{k}\hat Q^{\greater}_{\tS}
   -\eta_{\bar k}^{\ast}\hat Q^{\lesser}_{\tS})
  \rho^{(n);{\rm st}}_{{\bf n}_{kj}^{-+}},
  \\ &
   0=-\Big[iH^{\times}_{\tS}+\sum_k (n_k- \delta_{kj}) \gamma_{k}\Big]\rho^{(n-1);{\rm st}}_{{\bf n}^{-}_{j}}
  -i\sum_{k}\hat Q^{\times}_{\tS}\rho_{{\bf n}^{+-}_{kj}}^{(n);{\rm st}}
%\nl&
  -i\sum_{k}(n_{k}-\delta_{kj})(\eta_{k}\hat Q^{\greater}_{\tS}
   -\eta_{\bar k}^{\ast}\hat Q^{\lesser}_{\tS})
  \rho^{(n-2);{\rm st}}_{{\bf n}_{kj}^{--}}.
  \end{split}
\ee
Substituting \Eq{tmp3} into \Eq{tmp2}, followed by using the identity $\sum_k(\eta_k-\eta_{\bar k}^{\ast})=0 $ [cf. Eq.\,(9)] and some elementary algebra, we obtain
\be\label{lhs}
\dot{\rho}_{\bf n}^{(n)}(0;\hat x_{\B}^{\greater})
=\sum_k \gamma_k\big[
               \rho^{(n+1);{\rm st}}_{{\bf n}^{+}_{k}}
     -n_k\eta_k\rho^{(n-1);{\rm st}}_{{\bf n}^{-}_{k}}
   \big]
 =-\w_{\B}\rho^{(n)}_{{\bf n}}(0;\hat p_{\B}^{\greater}),
\ee
with the last identity being just \Eq{wick_p}.
We have thus proved that the proposed
dissipaton momentum algebra as Eq.\,(24)
does satisfy the sufficient criterion,
for the resulted phase--space DEOM evaluation on Eq.\,(3).

%\bibliographystyle{../aiptit}
%\bibliography{../bibrefs}